\documentclass[11pt]{article}
\usepackage{moriond,epsfig}

\def\Journal#1#2#3#4{{#1} {\bf #2}, #3 (#4)}

\def\NPO{{\em Nucl. Phys.}}

\def\PRL{\em Phys. Rev. Lett.}

\def\GeV{\ifmmode {\mathrm{\ Ge\kern -0.1em V}}\else
                   \textrm{Ge\kern -0.1em V}\fi}%
\def\ra{\rightarrow}
\def\HZ{\ensuremath{\mathrm{HZ}}}

\def\qq{\ensuremath{\mathrm{q\bar{q}}}}
\def\bb{\ensuremath{\mathrm{b\bar{b}}}}
\def\ee{\ensuremath{\mathrm{e^+e^-}}}

\def\tt{\ensuremath{\mathrm{\tau^+\tau^-}}}
\def\ll{\ensuremath{\mathrm{\ell^+\ell^-}}}
\def\nn{\ensuremath{\mathrm{\nu\bar{\nu}}}}
\def\WW{\ensuremath{\mathrm{W^+W^-}}}
\def\ZZ{\ensuremath{\mathrm{ZZ}}}

\def\pb{\mbox{pb$^{-1}$}}

\begin{document}
\vspace*{4cm}
\title{STANDARD MODEL HIGGS SEARCH STRATEGY AT LEP}

\author{P. GARCIA-ABIA}

\address{\footnote{On leave from the Institute of Physics of the University
         of Basel (Switzerland).}CERN, EP Division, 
         J21300, CH-1211-Geneva 23, Switzerland \\
         ~ \\
         On behalf of the LEP Collaborations
         }

\maketitle\abstracts{ 
The  Standard  Model Higgs boson has been  searched  for by the four LEP
experiments  in the last twelve years.  The data collected at LEP in the
year  2000  suggest  the first  observation  of a Higgs  boson.  In this
letter, I describe  the basic  concepts of the Higgs search at LEP, with
emphasis in the statistical  method used to combine the results from the
LEP experiments.}

\section{Introduction}

In 12 years of running,  the LEP  collider  at CERN has  produced  $\ee$
collisions  in the  interaction  region  of the  four  detectors  ALEPH,
DELPHI,  L3 and OPAL.  The  centre-of-mass  energy  of the  interactions
varied from about 90 to 210~\GeV.  The large amount of data collected at
these energies has allowed the  experiments to measure the most relevant
parameters  of the  Standard  Model~\cite{sm}  with very high  accuracy,
establishing the success of the model.

However, the Standard  Model does not provide an answer to a fundamental
problem:  in a gauge invariant theory all fundamental  particles  should
be massless, in contradiction  with the observation of the heavy W and Z
bosons.  Currently, there is no direct evidence of electroweak  symmetry
breaking and the  generation  of masses of the gauge  bosons.  The Higgs
mechanism is  introduced  in the model to let particles  acquire mass by
interaction  with the Higgs field.  As a result,  there  should be a new
particle of a  completely  new kind, the Higgs  boson, whose mass is not
predicted by the theory.

\section{Higgs Production at LEP}

\begin{figure}
\psfig{figure=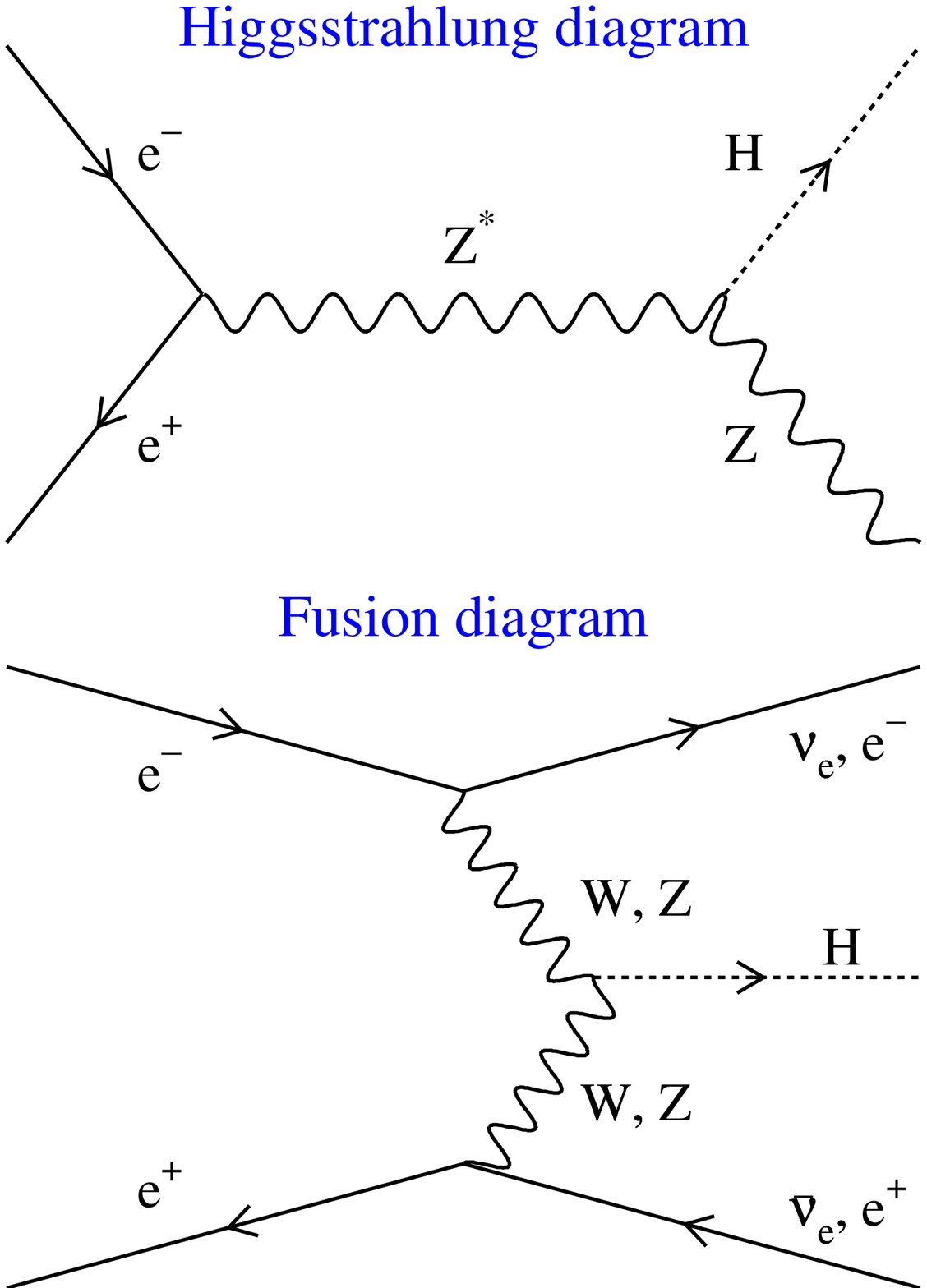,height=8cm}
\psfig{figure=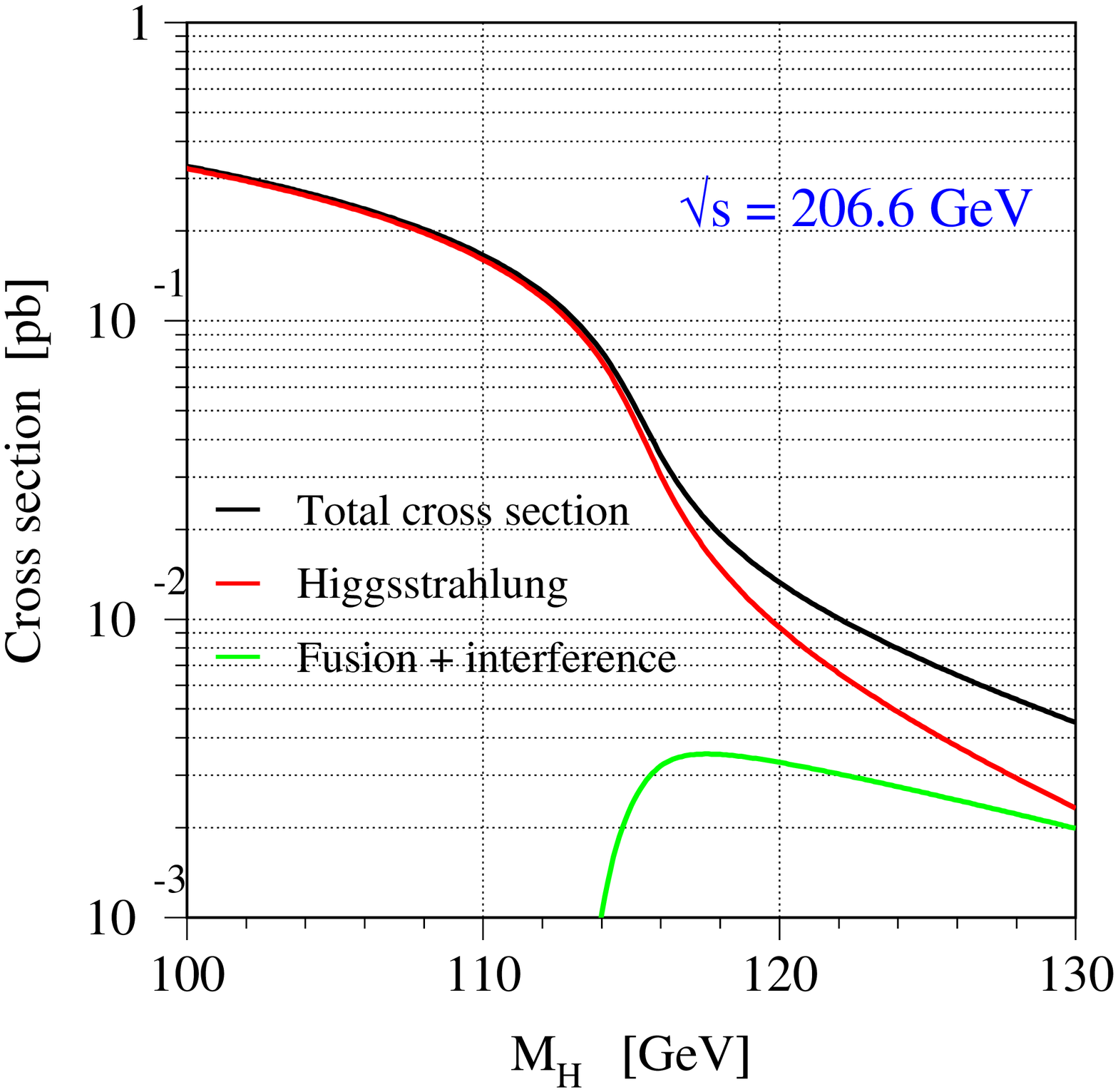,height=8cm}
\caption{Higgs  production modes (left) and cross section as function of
         the Higgs mass (right).
\label{fig:diags}}
\end{figure}

The dominant Higgs production mode at LEP is the Higgsstrahlung process,
$\ee\ra\mathrm{Z^\ast}\ra\HZ$   (Figure~\ref{fig:diags}   (left)).   The
processes $\WW$ and $\ZZ$ fusion, giving rise to the $\mathrm{H}\nn$ and
$\mathrm{H}\ee$  channels,  contribute  with  smaller  rate to the Higgs
production cross section (Figure~\ref{fig:diags} (right)).

The integrated  luminosity  collected by the four  experiments at LEP in
the year 2000 amounts to 870~\pb, at  centre-of-mass  energy between 200
and 209~\GeV.  At these  energies,  Higgs bosons of  116~\GeV{}  mass or
heavier are  difficult to be detected, as they would be produced  beyond
the  kinematic  limit.  As an  example,  assuming a Higgs of  115~\GeV{}
mass, a luminosity of 500~\pb  collected at $\sqrt{s} \sim 207~\GeV$ and
a detection efficiency of 50~\%, the number of expected signal events is
12.5.

The Higgs boson is predicted to decay mainly in b-quark pairs for masses
below 120~\GeV.  The Z boson decays into hadrons,  neutrinos and leptons
with  branching  fractions  69.6\%, 20\% and 10.1\%,  respectively.  The
search for the Higgs boson is based on the study of four distinct  event
topologies  representing  approximately 80\% of the $\HZ$ decay modes in
the mass range of interest:  $\bb\qq$ (4-jet channel), $\bb\nn$ (missing
energy  channel),   $\qq\ll\;(\ell=\mathrm{e},\mu,\tau)$   and  $\tt\qq$
(lepton  and tau  channels,  respectively).  With the  exception  of the
$\HZ\ra\tt\qq$   decay  mode,  all  the  analyses  are   optimised   for
$\mathrm{H}\ra\!\bb$ decay.

\begin{figure}                          \hspace{-0.5cm}
\psfig{figure=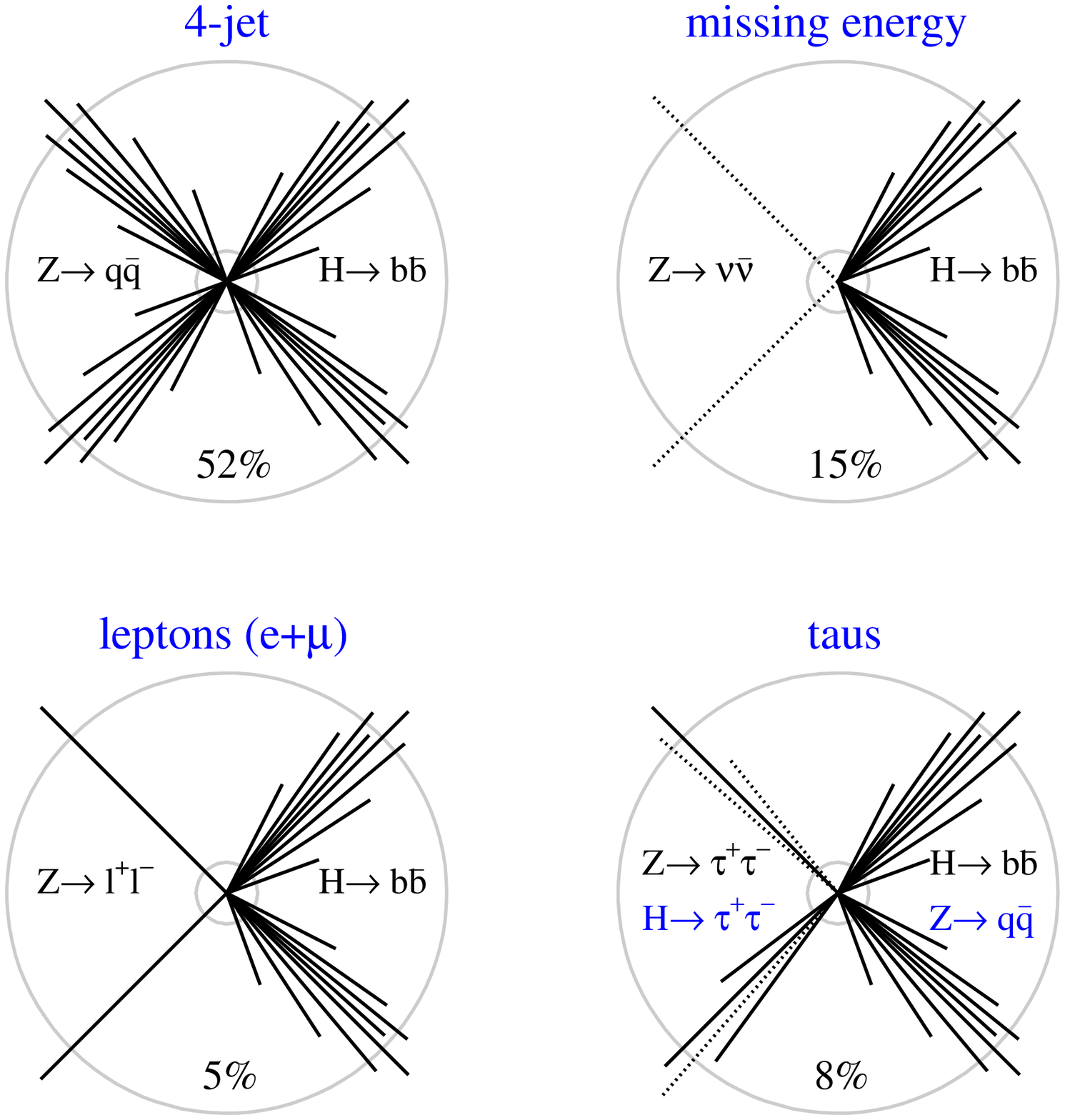,height=8.5cm}\hspace{0.5cm}
\psfig{figure=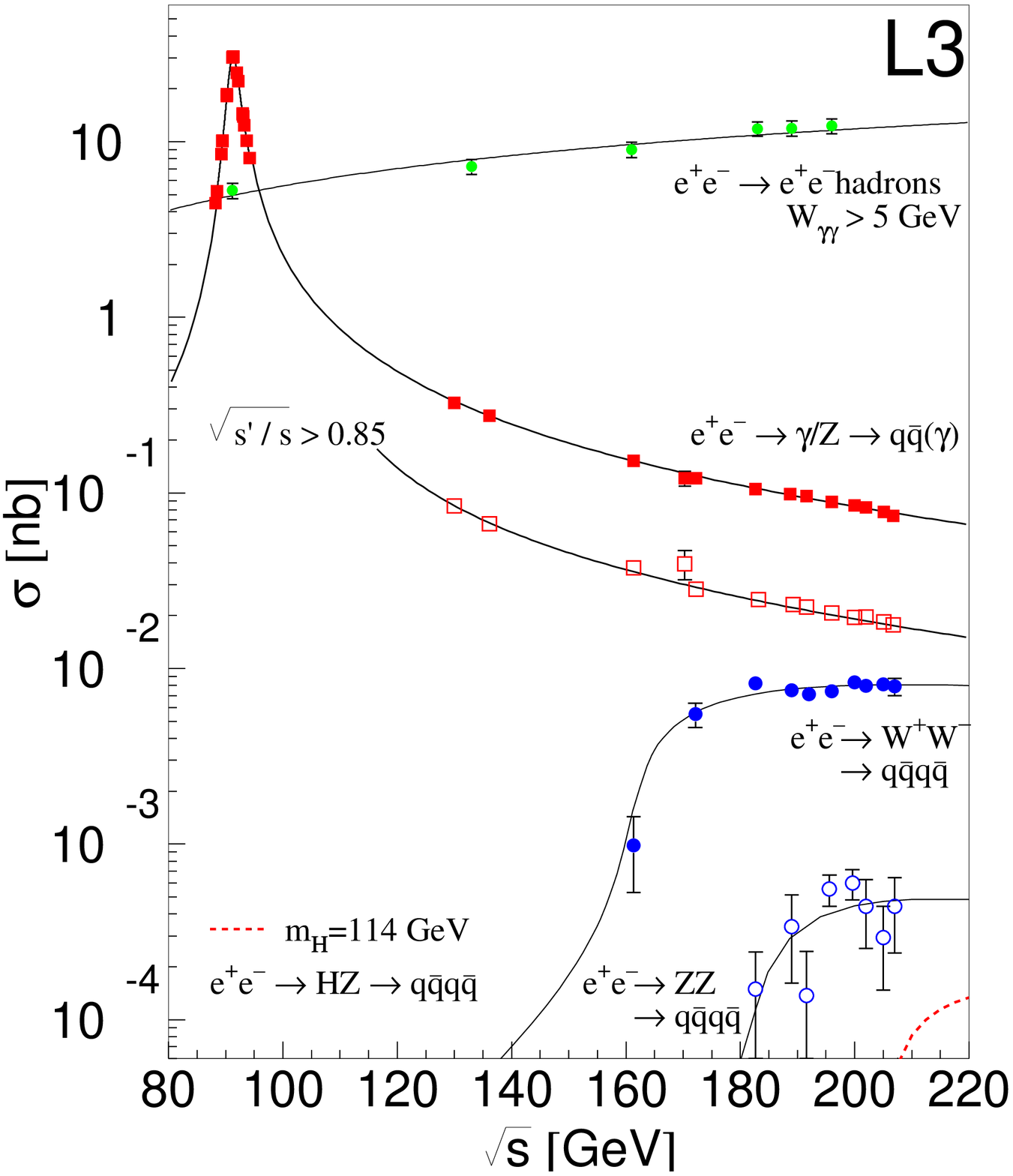,height=8.5cm}
\caption{Left,  topology of the four search channels,  together with the
         branching  ratio  and  the  conventional  name of the  channel.
         Right, cross section of the background processes.
\label{fig:topo}}
\end{figure}

Figure~\ref{fig:topo}  (left)  displays the  topology of the four search
channels,  together with their  branching  ratios.  The expected  signal
events for a 115~\GeV{} mass Higgs are:  6.5 from the 4-jet channel, 1.9
from the  missing  energy  channel,  1 from  taus  and 0.6 from  leptons
(e+$\mu$), that is 10 events in total.

The main  backgrounds  are events  from  $\ee\ra\qq(\gamma)$,  $\WW$ and
$\ZZ$  processes  (Figure~\ref{fig:topo}  (right)).  The tagging of jets
originating from b-quarks plays a crucial role in the  identification of
signal events.  After the final selections, typically few hundred events
are expected from background processes.  The detailed description of the
analyses  of the  different  channels  can be found  elsewhere  in these
proceedings~\cite{gd,ep}.

\mathversion{bold}
\section{The Statistical Method: $-2\,\mathrm{ln}(Q)$}
\mathversion{normal}

Each of the four LEP  experiments  performs  the  Higgs  search  in four
analysis  channels,  with  data  collected  at  several   centre-of-mass
energies.  A simple statistical  method~\cite{ll}  has been developed by
the experiments in order to treat LEP as a single search analysis.

The results of the individual analyses are expressed in terms of one (or
more)  final  variable,  also  called  {\em   discriminant},   which  is
calculated for the signal and background expectations as well as for the
data.  How this  final  variable  is  calculated  for  each  channel  is
explained   elsewhere~\cite{gd,ep}.  The  shape  of  the   distributions
depends on the Higgs mass hypothesis.

\begin{figure}
\psfig{figure=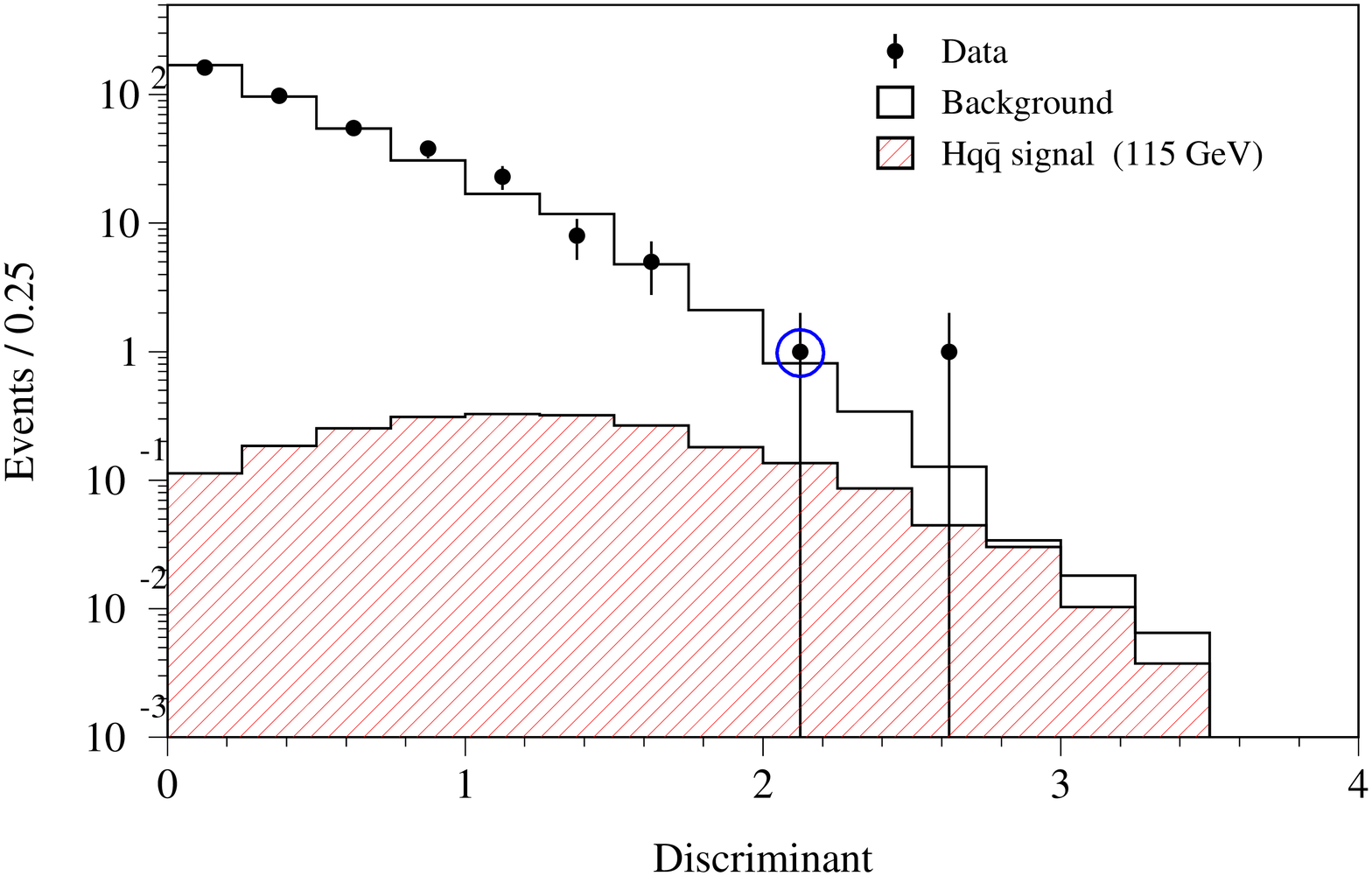,height=5.5cm}\hspace{0.5cm}
\psfig{figure=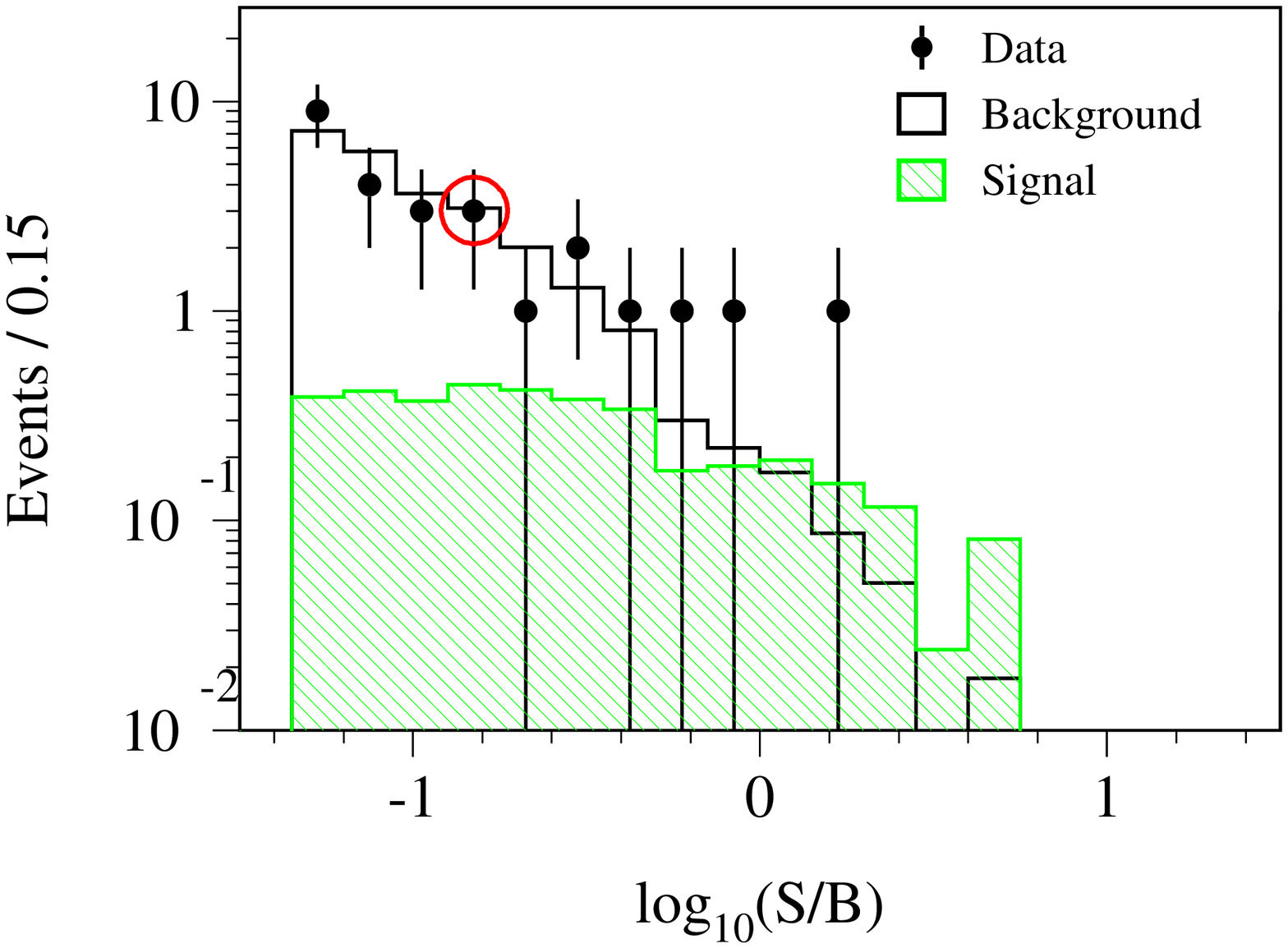,height=5.5cm}
\caption{Typical distributions of a discriminant variable (left) and the
         s/b ratio (right) for data, signal and background events.
\label{fig:disc}}
\end{figure}

The ratio between the expected  number of signal and  background  events
(the so  called  s/b  ratio)  for a given  value of the  final  variable
reflects the  likelihood  for an event being signal or  background-like.
Events from  different  experiments  and channels with the same value of
the s/b ratio are combined  together, as they have the same  sensitivity
(discriminating  power)  to a  possible  Higgs  signal.  As an  example,
Figure~\ref{fig:disc}  shows  typical  distributions  of a  discriminant
variable  (left)  and  the  s/b  ratio  (right)  for  data,  signal  and
background events.  Events at high values of s/b are more signal-like.

In  order  to  quantify  the  signal-likeness  of  the  data,  the  {\em
Likelihood  Ratio}~\cite{ll}  test-statistic  is defined as the ratio of
the  Poisson  probabilities  of data to be  consistent  with  either the
signal plus background hypothesis or the background only hypothesis:
$$Q = \frac{{\cal L}(s+b)}{{\cal L}(b)} $$
Each bin in the s/b  ratio  distribution  (i) is  treated  as a  Poisson
counting experiment:
$$\mathrm{ln}(Q) = -s_{tot} + \sum_{i=1}^N n_i \ \mathrm{ln}\left( 1 + \frac{s_i}{b_i}\right) $$
where $n_i$ is the number of observed  events, $s_i$ the expected signal
and $b_i$ the expected background in a given bin.  The sum runs over all
the bins in the s/b  distribution,  {\em i.e.}  all the observed  events
enter  the  likelihood  calculation.  The  value of $Q$, and  hence  its
discrimination  power, depends on the Higgs mass  hypothesis  and on the
Higgs  production  cross  section  ($s_{tot}$).  In the high  statistics
limit $-2\,\mathrm{ln}(Q)$  approaches the difference in $\chi^2$ of the
two hypotheses, that is why this convention is adopted.

\begin{figure}
\psfig{figure=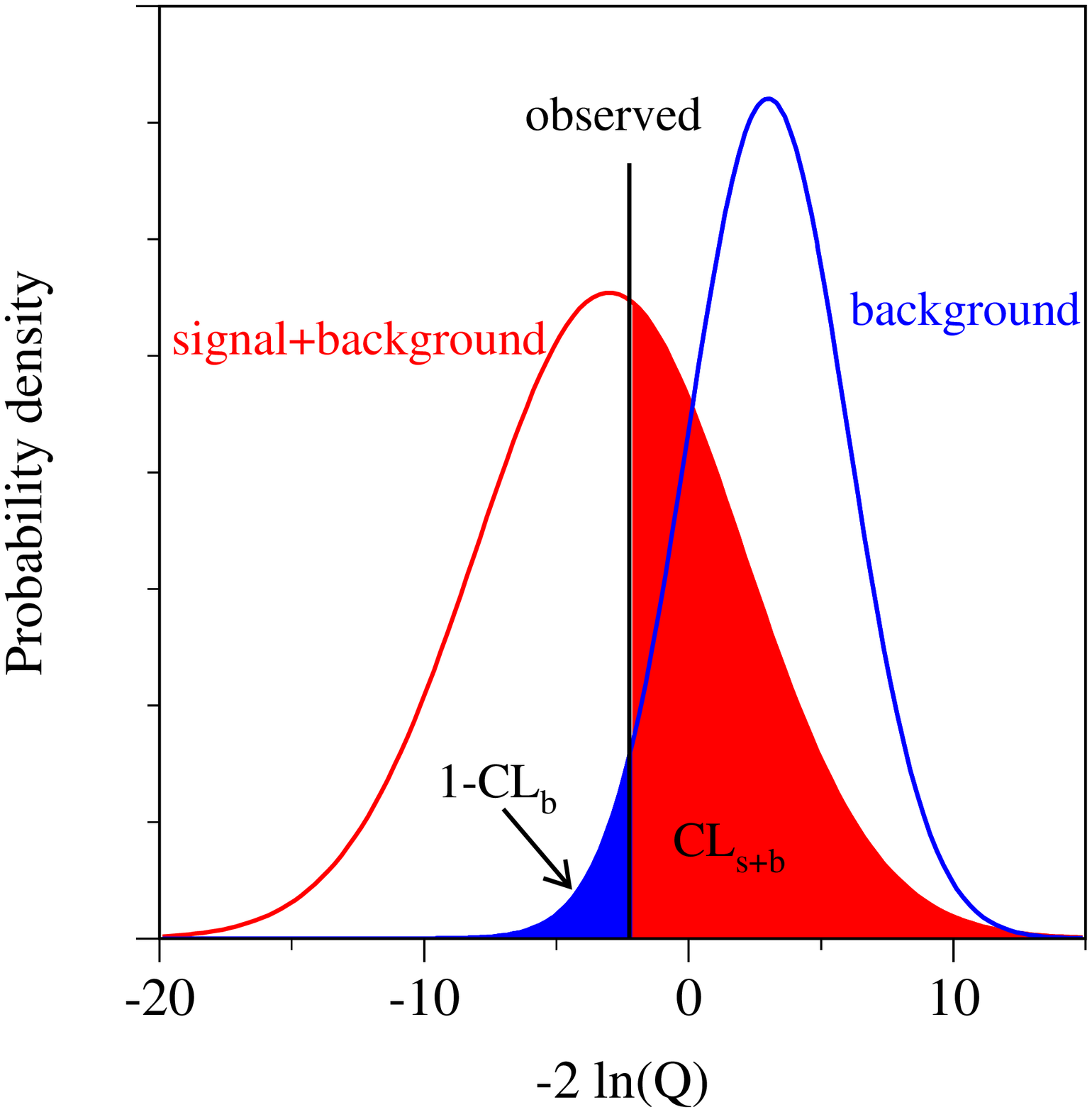,height=7.5cm}\hspace{0.2cm}
\psfig{figure=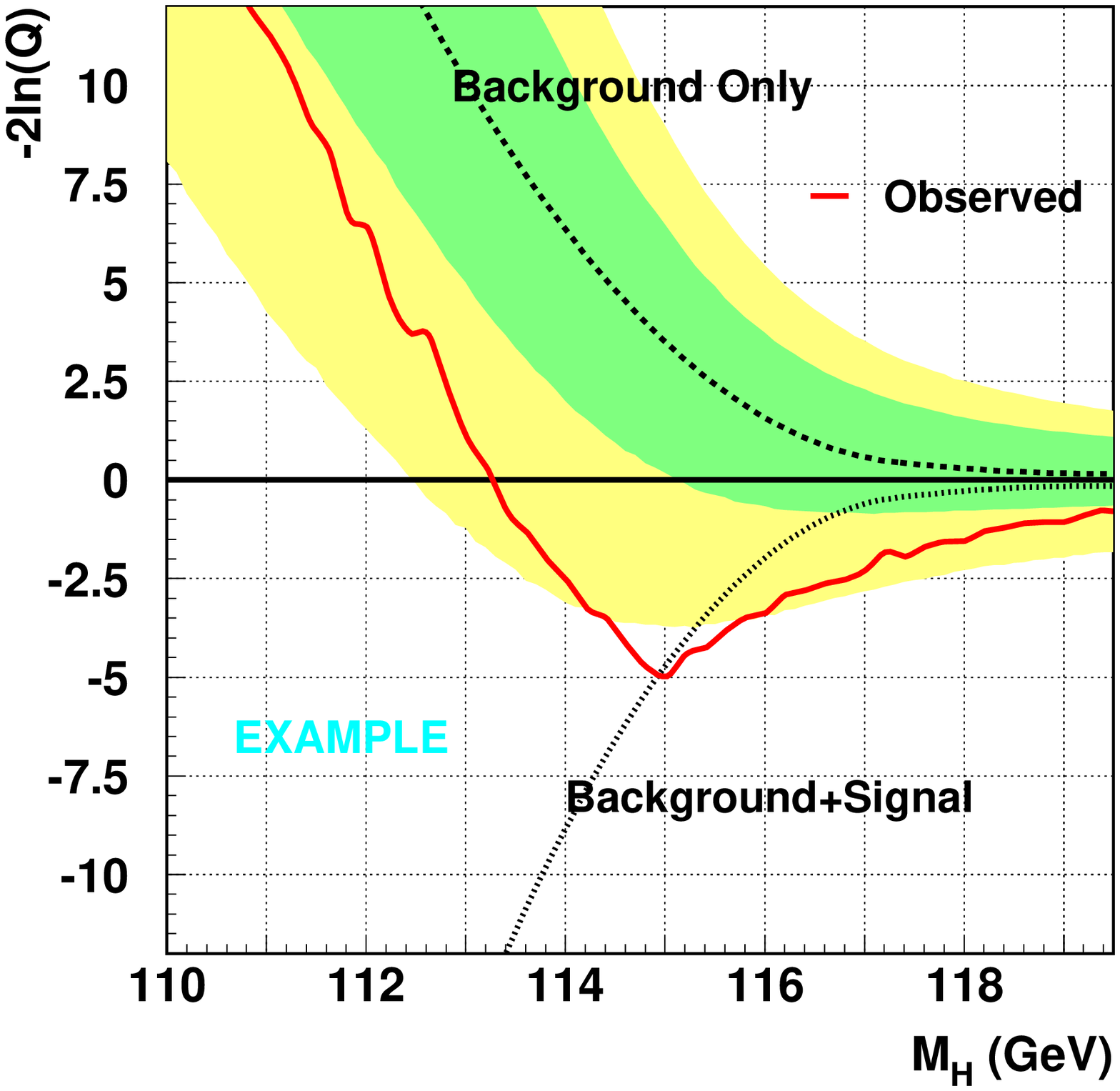,height=7.5cm}
\caption{Left,  example  distributions of  $-2\,\mathrm{ln}(Q)$  for the
         background and signal plus  background  hypotheses,  as well as
         for the data, for a given  Higgs mass.  The  definition  of the
         confidence levels of the background ($\mathrm{{CL}_b}$) and the
         signal plus background  ($\mathrm{{CL}_{s+b}}$) are also shown.
         Right, example distribution of $-2\,\mathrm{ln}(Q)$ as function
         of the Higgs  mass.  The  colour  bands are the  $1\sigma$  and
         $2\sigma$ bands for the background expectation.
\label{fig:lnq}}
\end{figure}

The results of the LEP  combination  using this  statistical  method are
presented in detail elsewhere in these proceedings~\cite{ao}.

\end{document}